\documentclass[twoside,twocolumn,9pt]{article}	
\usepackage{extsizes}
\usepackage[super,sort&compress,comma]{natbib} 
\usepackage[version=3]{mhchem}
\usepackage[left=1.5cm, right=1.5cm, top=1.785cm, bottom=2.0cm]{geometry}
\usepackage{balance}
\usepackage{mathptmx}
\usepackage{sectsty}
\usepackage{graphicx} 
\usepackage{lastpage}
\usepackage[format=plain,justification=justified,singlelinecheck=false,font={stretch=1.125,small,sf},labelfont=bf,labelsep=space]{caption}
\usepackage{fancyhdr}
\usepackage{fnpos}
\usepackage[english]{babel}
\addto{\captionsenglish}{%
  
}
\usepackage{array}
\usepackage{droidsans}
\usepackage{charter}
\usepackage[T1]{fontenc}
\usepackage[usenames,dvipsnames]{xcolor}
\usepackage{setspace}
\usepackage[compact]{titlesec}
\usepackage{hyperref}

\usepackage{epstopdf}

\definecolor{cream}{RGB}{222,217,201}

\usepackage{amssymb}
\usepackage{amsmath}
\usepackage{amsbsy}
\usepackage{xcolor}
\usepackage{soul}
\usepackage{array}
\usepackage{appendix}
\usepackage{float}

\begin{document}

\pagestyle{fancy}
\thispagestyle{plain}
\fancypagestyle{plain}{
\renewcommand{\headrulewidth}{0pt}
}

\makeFNbottom
\makeatletter
\renewcommand\LARGE{\@setfontsize\LARGE{15pt}{17}}
\renewcommand\Large{\@setfontsize\Large{12pt}{14}}
\renewcommand\large{\@setfontsize\large{10pt}{12}}
\renewcommand\footnotesize{\@setfontsize\footnotesize{7pt}{10}}
\makeatother

\renewcommand{\thefootnote}{\fnsymbol{footnote}}
\renewcommand\footnoterule{\vspace*{1pt}%
\color{cream}\hrule width 3.5in height 0.4pt \color{black}\vspace*{5pt}} 
\setcounter{secnumdepth}{5}

\makeatletter 
\renewcommand\@biblabel[1]{#1}            
\renewcommand\@makefntext[1]%
{\noindent\makebox[0pt][r]{\@thefnmark\,}#1}
\makeatother 
\renewcommand{\figurename}{\small{Fig.}~}
\sectionfont{\sffamily\Large}
\subsectionfont{\normalsize}
\subsubsectionfont{\bf}
\setstretch{1.125} 
\setlength{\skip\footins}{0.8cm}
\setlength{\footnotesep}{0.25cm}
\setlength{\jot}{10pt}
\titlespacing*{\section}{0pt}{4pt}{4pt}
\titlespacing*{\subsection}{0pt}{15pt}{1pt}

\fancyfoot{}
\fancyfoot[LO,RE]{\vspace{-7.1pt}\includegraphics[height=9pt]{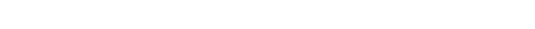}}
\fancyfoot[CO]{\vspace{-7.1pt}\hspace{13.2cm}\includegraphics{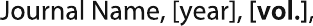}}
\fancyfoot[CE]{\vspace{-7.2pt}\hspace{-14.2cm}\includegraphics{head_foot/RF}}
\fancyfoot[RO]{\footnotesize{\sffamily{1--\pageref{LastPage} ~\textbar  \hspace{2pt}\thepage}}}
\fancyfoot[LE]{\footnotesize{\sffamily{\thepage~\textbar\hspace{3.45cm} 1--\pageref{LastPage}}}}
\fancyhead{}
\renewcommand{\headrulewidth}{0pt} 
\renewcommand{\footrulewidth}{0pt}
\setlength{\arrayrulewidth}{1pt}
\setlength{\columnsep}{6.5mm}
\setlength\bibsep{1pt}

\makeatletter 
\newlength{\figrulesep} 
\setlength{\figrulesep}{0.5\textfloatsep} 

\newcommand{\topfigrule}{\vspace*{-1pt}%
\noindent{\color{cream}\rule[-\figrulesep]{\columnwidth}{1.5pt}} }

\newcommand{\botfigrule}{\vspace*{-2pt}%
\noindent{\color{cream}\rule[\figrulesep]{\columnwidth}{1.5pt}} }

\newcommand{\dblfigrule}{\vspace*{-1pt}%
\noindent{\color{cream}\rule[-\figrulesep]{\textwidth}{1.5pt}} }

\makeatother

\twocolumn[
  \begin{@twocolumnfalse}
{\includegraphics[height=30pt]{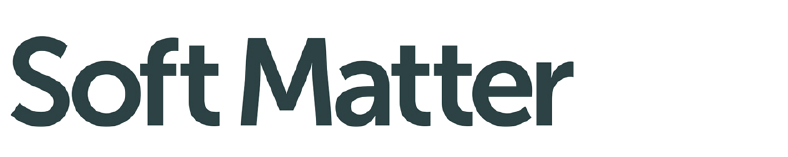}\hfill\raisebox{0pt}[0pt][0pt]{\includegraphics[height=55pt]{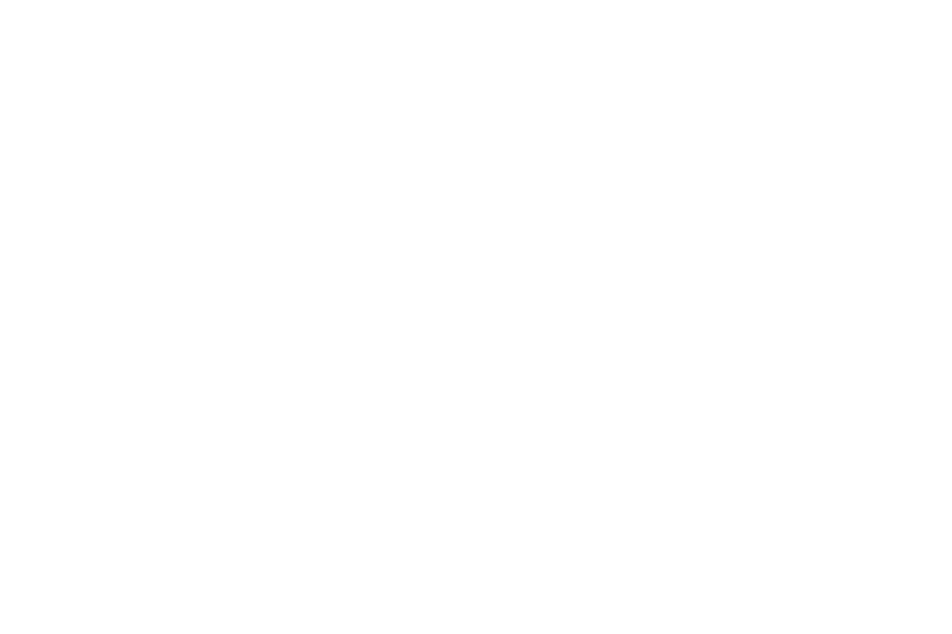}}\\[1ex]
\includegraphics[width=18.5cm]{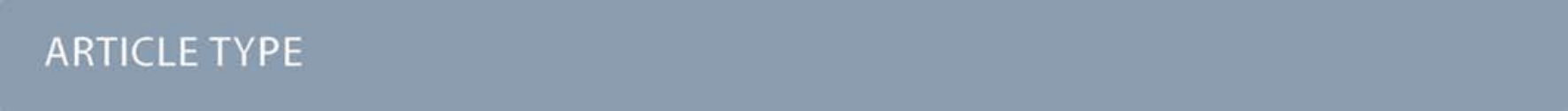}}\par
\vspace{1em}
\sffamily
\begin{tabular}{m{4.5cm} p{13.5cm} }

\includegraphics{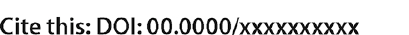} & \noindent\LARGE{\textbf{Viscoelasticity and elastocapillarity effects in the impact of drops on a repellent surface$^\dag$}} \\
\vspace{0.3cm} & \vspace{0.3cm} \\

 & \noindent\large{Carole-Ann Charles,\textit{$^{a}$} Ameur Louhichi,\textit{$^{a}$} Laurence Ramos, \textit{$^{a}$} and Christian Ligoure \textit{$^\ast$} \textit{$^{a}$}} \\

\includegraphics{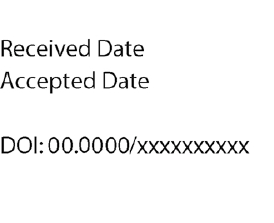} & \noindent\normalsize{We investigate freely expanding viscoelastic sheets. The sheets are produced by the impact of drops on a quartz plate covered with a thin layer of liquid nitrogen that suppresses shear viscous dissipation as a result of the cold Leidenfrost effect. The time evolution of the sheet is simultaneously recorded from top and side views using high-speed cameras. The investigated viscoelastic fluids are Maxwell fluids, which are characterized by low elastic moduli, and relaxation times that vary over almost two orders of magnitude, thus giving access to a large spectrum of viscoelastic and elastocapillary effects. For the purposes of comparison, Newtonian fluids, with viscosity varying over three orders of magnitude, are also investigated. In this study, $d_{\mathrm{max}}$, the maximal expansion of the sheets, and  $t_{\mathrm{max}}$  the time to reach this maximal expansion from the time at impact, are measured as a function of the impact velocity. By using a generalized damped harmonic oscillator model, we rationalize the role of capillarity, bulk elasticity and viscous dissipation in the expansion dynamics of all investigated samples. In the model, the spring constant is a combination of the surface tension and the bulk dynamic elastic modulus. The time-varying damping coefficient is associated to biaxial extensional viscous dissipation and is proportional to the dynamic loss modulus. For all samples, we find that the model reproduces accurately the experimental data for $d_{\mathrm{max}}$ and $t_{\mathrm{max}}$.
} \\

\end{tabular}

 \end{@twocolumnfalse} \vspace{0.6cm}

 ]

\renewcommand*\rmdefault{bch}\normalfont\upshape
\rmfamily
\section*{}
\vspace{-1cm}


\footnotetext{\textit{$^{a}$~ Laboratoire Charles Coulomb (L2C), Universit\'e de  Montpellier, CNRS, Montpellier, France}}

\footnotetext{\dag~Electronic Supplementary Information (ESI) available: [details of any supplementary information available should be included here]. See DOI: 10.1039/cXsm00000x/}

\footnotetext{\textit{$^\ast$} Corresponding author: christian.ligoure@umontpellier.fr}


\section{Introduction}
	
When a drop with a high level of kinetic energy hits a surface, it expands radially into a transient sheet, until reaching a maximum diameter before, in some cases, receding partially or completely depending on the nature of the drop and the surface. Predicting the maximum diameter and the time needed for the sheet to reach the latter is crucial for many industrial applications such as spray coating~\cite{Pasandideh-Fard2002}, pesticide application \cite{Wirth1991}, forensic science \cite{Laan2015} or ink-jet printing~\cite{Wijshoff2018}. As a matter of fact, the dynamics of drop impact has been extensively studied over the years and has been greatly aided by the emergence of high speed imaging~\cite{Thoroddsen2008, Josserand2016}. The impact of a drop can now be observed in real time for various systems such as Newtonian fluids with different viscosities~\cite{Wang2017, Gordillo2018}, suspensions of particles \cite{Marston2013,Raux:2020hs}, shear thickening fluids~\cite{Boyer2016}, polymer solutions~\cite{Crooks2000} or other viscoelastic materials~\cite{Arora2016} as well as soft elastic beads~\cite{Arora2018}. These materials are impacted on many different surfaces such as thick and thin liquid films or solid surfaces of different roughness~\cite{Rioboo:2002gk} and wettability~\cite{Lee:2010bt} but also of different sizes, geometries~\cite{Josserand2016, Wang2017, Arogeti2019} or inclinations~\cite{Moreira:2007gh}.\\
	Numerous rationalizations \cite{Josserand2016} and simulations \cite{YWAng2019,YWang2020,XWang2020} of the impact process have provided great insight into the expansion dynamics of Newtonian fluids. These models are generally based on a balance of energy and aim to predict the maximum diameter reached by the sheet as a function of characteristic adimensional numbers (Reynolds, Weber and Ohnesorge numbers). In the case of viscoelastic fluids, rationalization is however more complex and was not as extensively studied~\cite{Bertola2013,shah2020drop}. Though, because most industrial impact phenomenon involve viscoelastic fluids, understanding the behavior of impacted viscoelastic drops is of great interest.\\ 
Impact dynamics of viscoelastic fluids is more complex than for Newtonian fluids. For this reason, we simplify the contribution of dissipation by impacting drops on a repellent surface. Such surfaces include superhydrophobic surfaces~\cite{Richard2002}, hot plates above the Leidenfrost temperature~\cite{Wachters:1966vy, Biance2003} or cold plates used in cold Leidenfrost conditions~\cite{Antonini2013, Arora2018}. Repellent surfaces avoid a direct contact between the liquid sheet and the solid surface hence it suppresses the contact between the drop and the solid surface during drop collision and sheet expansion. Viscous dissipation due to shear is therefore suppressed. However, the use of repellent surfaces does not suppress all viscous dissipation. We have recently shown that biaxial extensional viscosity is the relevant source of viscous dissipation involved in the impact dynamics of drops in cold Leidenfrost conditions for which the shear dissipation is suppressed~\cite{Louhichi2020}.\\
	In this work, we investigate the expansion dynamics of freely expanding sheets formed by the impact of drops of viscoelastic fluids for which the effects of bulk elasticity, capillarity and viscosity combine in a non trivial way. Drops are impacted in cold Leidenfrost conditions at several impact velocities. A simple model of 1D harmonic oscillator was used to provide a quantitative estimate of the expansion dynamics of sheets without viscous dissipation for drops of inviscid fluids~\cite{Biance2006, Andrew2017}, for elastic beads~\cite{Tanaka:2006gp} and for ultra soft elastic beads and viscoelastic drops with long relaxation times with respect to the duration of the impact experiment \cite{Arora2018}. Viscous damping has also been recently included in the 1D oscillator model to account for viscous dissipation in the rebound of Newtonian fluids \cite{Jha2020} or droplet oscillation after impact \cite{Lin2018}. Here, we show that taking into account the viscoelastic nature of the samples is crucial to capture into a generalized damped oscillator model, beyond the scaling arguments, the non-trivial combination of bulk elasticity, surface tension and viscosity in the expansion dynamics of viscoelastic drops following impact. The model successfully describes the expansion of sheets made from viscoelastic fluids as well as Newtonian liquids with a wide range of viscosities allowing one to extend investigations further than the capillary regime.\\

\section{Methods and materials}
	
\subsection{Experimental set-up}
	
	 	With the aim of weeding out the shear dissipation, drop impact experiments are performed under cold Leidenfrost conditions using a set-up described previously~\cite{Arora2018,Louhichi2020}. The drop at ambient temperature impacts a quartz plate covered by a thin layer of liquid nitrogen (boiling point $T_\textrm{s}$=-196.15$^\circ$C). As the temperature of the drop is higher than $T_\textrm{s}$, the liquid nitrogen evaporates partially upon impact, creating a vapor layer under the expanding drop of typical thickness $100 \mu$m \cite{Quere2013}. The drop is supported by the evaporating nitrogen throughout the whole expansion process and is left to expand free of shear viscous dissipation~\cite{Antonini2013,Chen:2016fo}.
		
\begin{figure}[h]
	\includegraphics[width=\columnwidth]{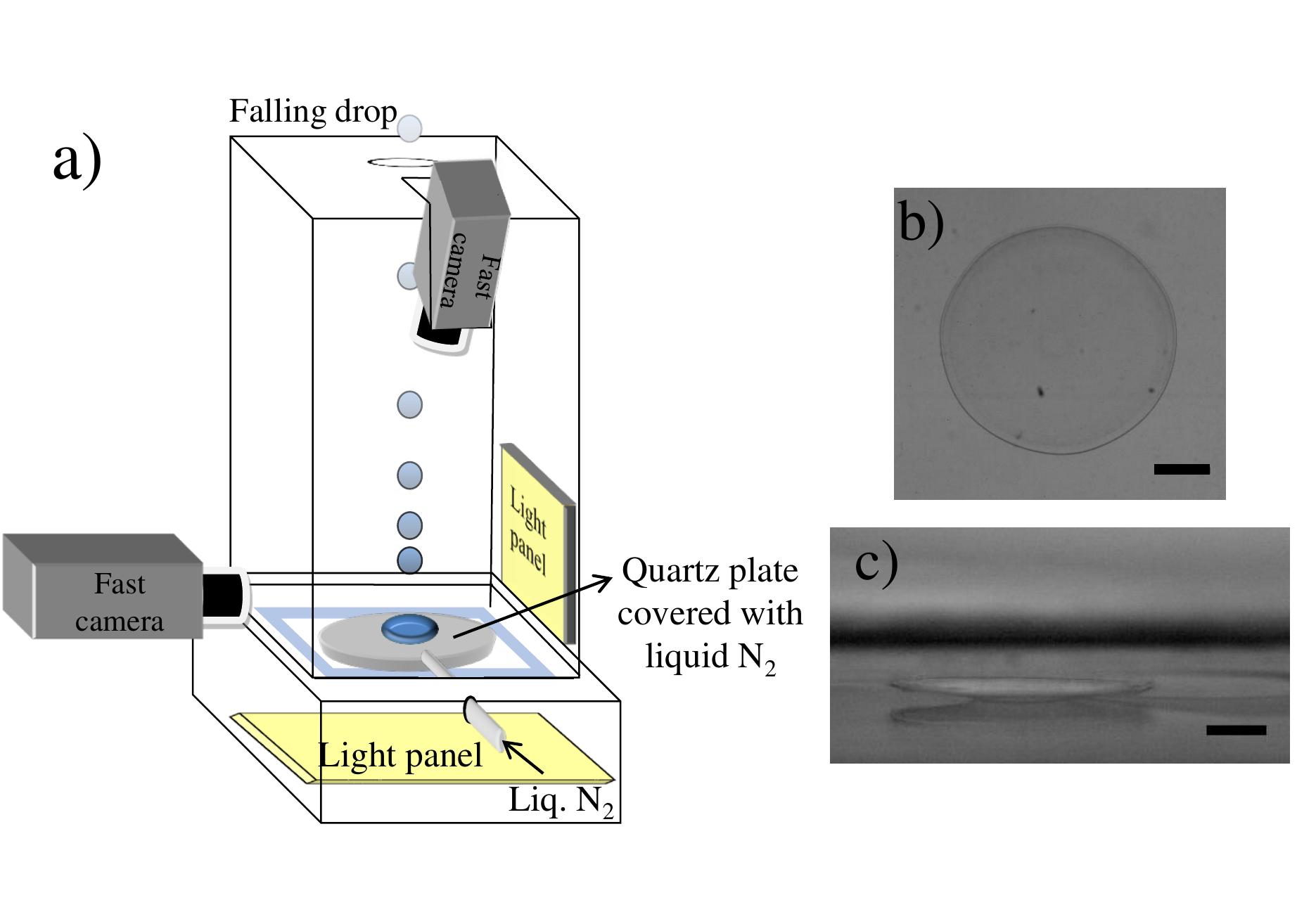}
	\caption{ (a) Schematic illustration of the set-up where a drop falls onto a quartz plate covered with a thin layer of liquid nitrogen. The impact is simultaneously recorded from the top and the side by two fast cameras. (b,c) Snapshots of a viscoelastic sheet (sample M14$\phi$10\textit{r}6) taken at maximal
 expansion from the top (b) and from the side (c). The impact velocity is $v_0=4.7$ m/s. The scale bars represent $5$ mm.}
	\label{fig1}
	\end{figure}

The experimental set-up is schematically shown in figure~\ref{fig1}a. The drops are released from a needle (internal diameter 2 mm) placed vertically above the quartz surface and connected, through a flexible Teflon tube, to a syringe pump set at a flow rate of 1 ml/min. The initial diameter of the drop, $d_0$, varies between $3.1$ and $3.8$ mm depending on the fluids surface tension. The height, $H$, from which the drop is released into a free fall is varied between $11$ cm and $131$ cm in order to obtain impact velocities ranging between $1.5$ and $5$ m/s. Drops accumulate inertial energy throughout their fall until reaching the surface covered with the thin liquid nitrogen layer where they expand and reach a maximal expansion before retracting due to surface tension and stored elastic energy. During retraction, some droplets are eventually expelled from the rim. The whole process is captured from the top, at an angle of approximately $10^\circ$ with the vertical plane, by a high speed camera, Phantom V 7:3, operating at $6700$ fps with a resolution of $800 \times 600$ pixels$^2$ and, for most experiments, simultaneously from the side by a Phantom miro M310 operating at $3200$ fps with a resolution of $1280 \times 800 $ pixels$^2$. We use the side imaging to check that the sheets are flat during the whole duration of the expansion. The top view imaging is used to quantify the expansion dynamic. Proper illumination for acquisition is provided by high-intensity backlights; Phlox HSC with a luminance of $98$ cd/m$^2$ in the bottom and Phlox LLUB with a luminance of $20$ cd/m$^2$ on the side. The temperature (20$^\circ $C on average) of the room is systematically measured before each impact experiment to control the experimental conditions. Between each impact, the substrate is cleaned with ethanol. Figures~\ref{fig1}b,c show top and side views of a viscoelastic sheet taken at its maximal expansion. The snapshots show that the sheet at maximal expansion has a near circular shape and lays almost flat on the nitrogen gas.

		As the temperature of the fluid on which the drop is impacted is very low ($T_\textrm{s}=-196.15^\circ$C, the boiling point of liquid nitrogen) as compared to the initial temperature of the drop (about $20^\circ$C), one should ensure that the drop does not freeze upon impact and that its temperature remains constant throughout the expansion. We thus quantify the typical time for the heat to transfer from the drop to the quartz plate and compare this time to the typical time of an experiment ( typically of the order of a few ms). The principal heat transfer mode between the sample and the vapor layer is conduction \cite{Hendricks1970,Hendricks1971,Kim2011}. The calculation is reported in the supplementary materials \dag. It shows that the temperature of the sheet decreases by less than $0.8^\circ$C when the sheet reaches maximal expansion. Hence, it is reasonable to assume that the temperature of the sheet does not vary considerably until maximum expansion. Thus, in the following the rationalization of the experimental results will be done using the fluids properties measured at $20^\circ $C, close to the temperature of the drop impact experiments.

	\subsection{Image analysis}
	
	The stacks of frames obtained from top view imaging are analyzed using ImageJ software. The sheet is delimited by adjusting the threshold and its area, \textit{A}, is measured for each frame using the analyze particle command. We have checked that corrections in the measured area, due to the camera angle, can be safely neglected. An apparent diameter, $d$, is extracted using $d=\sqrt{\frac{4A}{\pi}}$. For Newtonian fluids with zero shear viscosities lower than $100$ mPa s, the maximal diameter is corrected in accordance with observations made in Ref.\cite{Louhichi2020}. Indeed, for these low viscosity samples side view imaging shows that the sheet exhibits a corona shape, resulting in an underestimation of the actual diameter if simply estimated from top imaging \cite{Louhichi2020}. We check that for all investigated viscoelastic samples, the sheets remain flat and no correction is necessary. In the following, for each impact velocity and sample, the reported time evolution of the sheet diameter corresponds to an average over three different experiments.\\

	\subsection{Shear rheology}
	
	All samples are characterized rheologically using either Anton Paar MRC302 or ARES RFS 1KFRT rheometers, equipped with stainless steel cone and plate geometries of diameter $50$ mm with an angle of 1$^\circ$ and a truncation of 101 $\mu$m. Dynamic strain sweeps with strain amplitude from $0.1$\% to $100$\% are conducted at frequencies of $10$ and $50$ rad/s to define the linear viscoelastic regime. Dynamic frequency sweeps are performed at a strain amplitude $\gamma=1\%$, well in the linear regime, for an angular frequency varying from $100$ to $0.01$ rad/s. Temperature is set at $20^\circ$C for all measurements and controlled using a Peltier element (accuracy of $0.2^\circ$C).\\

	\subsection{Systems}
	
	Two classes of fluids are investigated: two types of Newtonian liquids and two types of viscoelastic fluids. 

The Newtonian liquids consist of silicone oil blends (SO) and glycerol-water mixtures (G-W). Silicone oils are purchased from Sigma Aldrich and blended to give mixtures of different viscosities. The zero shear viscosities of the blends vary from $5$ mPa s to $970$ mPa s as measured at $20^\circ$C, their average surface tension is $20$ mN/m~\cite{Crisp1987} and their density varies between $0.913$ and $0.97$ g/ml (as calculated from the densities of the purchased silicone oils). The composition of glycerol-water mixtures varies from $0$ to $97.5$ \% w/w glycerol, yielding zero shear viscosities from $1$ mPa s (for pure water (W)) to $813$ mPa s (for $97.5$ \% w/w glycerol), densities from $1.05$ g/ml to $1.25$ g/ml, and an average surface tension of $65$ mN/m, as measured with a pendant drop set-up.

	The first viscoelastic system consists of bridged micro-emulsions composed of decane droplets of radius $6$ nm~\cite{Filali2001,Arora2018} stabilized in brine ($0.2$ M NaCl) by cetylpyridinium chloride (CpCl) as surfactant, and n-octanol as co-surfactant, with a molar ratio of n-octanol over CpCl of $0.65$. The droplets are reversibly linked by polymer chains of poly(ethylene oxide) (PEO) of molar mass $35$ kg/mol with hydrophobic stickers grafted at both ends, resulting in the formation of a transient network. The micro-emulsions are characterized by the number of carbon atoms in each sticker, $n$ ($n=12$, $14$ or $18$), the average number of stickers per droplet, $r$, comprised between $4$ and $9$, and $\phi$, the mass fraction of oil droplets ($\phi=8, 10$ \%). The sample preparation is described elsewhere~\cite{Arora2016}.

	The second viscoelastic system consists of entangled wormlike micelles \cite{Rehage1988,Rehage1991} produced by self-organization, in $0.5$ M brine, of surfactant molecules, sodium salicylate (NaSal) and CpCl, with a fixed molar ratio [NaSal]/[CpCl] of $0.5$. Micelles are decorated by an amphiphilic polymer ($Synperonic^{®}$ F-108), purchased from Serva, a PEO-PPO-PEO (with PPO being polypropylene oxide) triblock copolymer with an average molar mass of $14$ kg/mol. The addition of amphiphilic polymer allows to easily tune the relaxation time \cite{massi2002}. The micelles preparation is described in ref.\cite{Massiera2002a}. Micelles are characterized by $\phi$, the mass fraction of surfactant (between $5$ and $9$\%) and $\alpha$ the molar ratio of amphiphilic polymer over surfactant (from 0.48 to 4$\%$).

\begin{table}[!ht]
\centering
	\small
		\caption{List of viscoelastic samples and their rheological properties. Micro-emulsion samples are named M$n\phi$X\textit{r}Y with $n$, the number of carbon per sticker for the telechelic polymers, X the value in $\%$ g/g for $\phi$, the hydrophobic weight fraction and Y, the value of $r$, the average number of stickers per oil droplet. The name of wormlike micelle samples follows the nomenclature: WM$\phi$X$\alpha$Y with X the value in $\%$ g/g for $\phi$, the mass fraction of surfactant, and Y the value in $\%$ for $\alpha$, the mole fraction of amphiphilic polymer. $G_0$ is the elastic modulus, $\tau$, the relaxation time and $\eta_0=G_0\tau$, the zero shear viscosity.}
		\label{t}
		\begin{tabular*}{0.48\textwidth}{@{\extracolsep{\fill}}llll}
		\hline
		\multicolumn{4}{l}{Micro-emulsions} \\
		\hline
		Name & $G_\textrm{0}$ [Pa] & $\tau$ [ms]& $\eta_\textrm{0}$ [Pa s] \\
		\hline
		M18$\phi$10\textit{r}4 & 10 & 178 & 1.78 \\
		\hline
		M14$\phi$8\textit{r}9 & 189 & 8 & 1.5 \\
		\hline
		M14$\phi$8\textit{r}8 & 128 & 6 & 0.77 \\
		\hline
		M14$\phi$10\textit{r}6 & 48 & 5 & 0.23 \\
		\hline
		M14$\phi$8\textit{r}6 & 31 & 4 & 0.12 \\
		\hline
		M12$\phi$10\textit{r}8 & 194 & 2 & 0.39 \\
		\hline
		\multicolumn{4}{l}{Wormlike micelles} \\
		\hline
		Name & $G_\textrm{0}$ [Pa] & $\tau$ [ms]& $\eta_\textrm{0}$ [Pa s] \\
		\hline
		WM$\phi$5$\alpha$0.48 & 50 & 8 & 0.4 \\
		\hline
		WM$\phi$7$\alpha$1.8 & 73 & 2 & 0.146 \\
		\hline
		WM$\phi$9$\alpha$4 & 64 & 1 & 0.031 \\
		\hline
		\end{tabular*}

		 \end{table}

\begin{figure}[!ht]
	
	\includegraphics[width=8.3 cm]{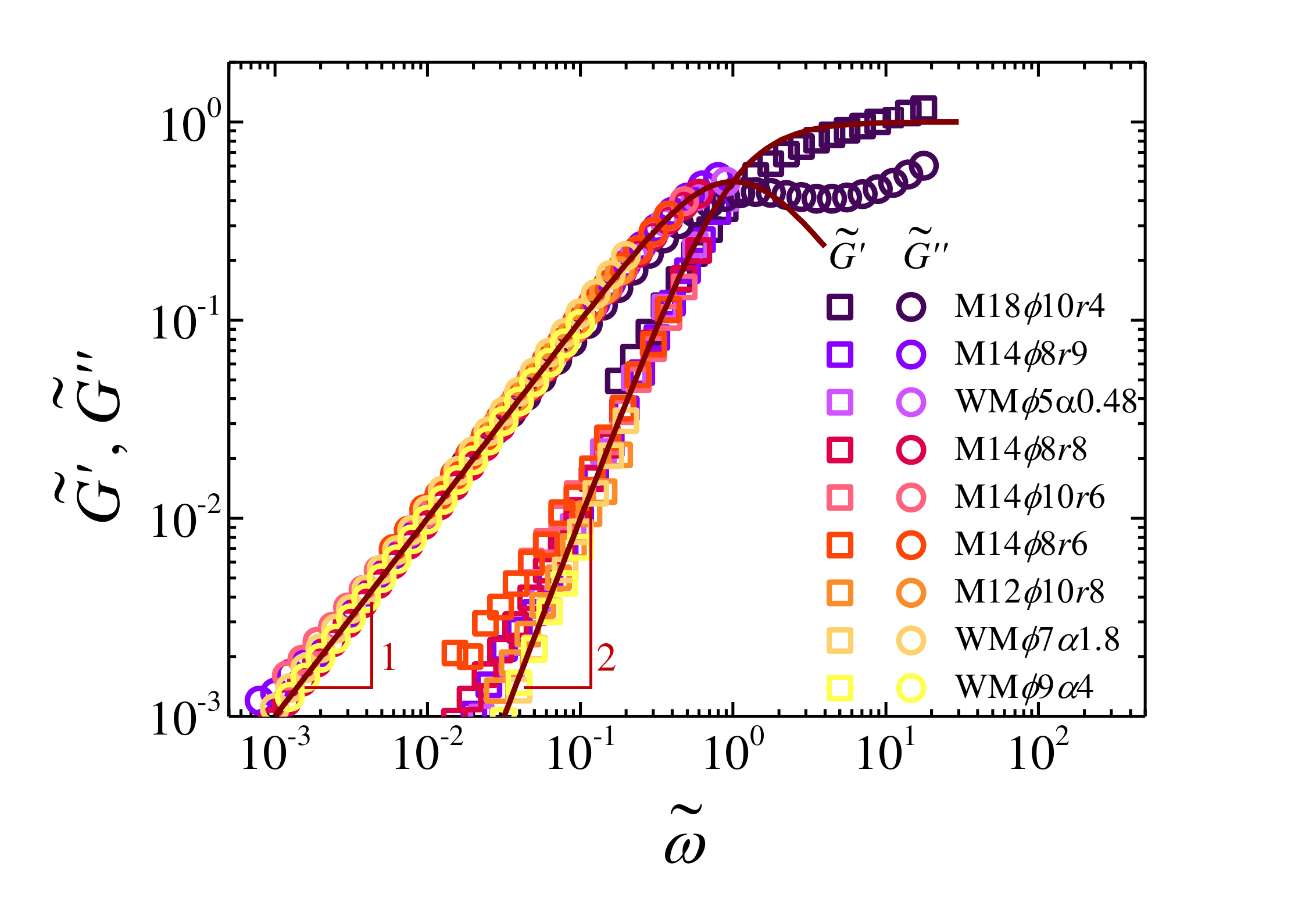}
	
	\caption{(Symbols) Reduced storage, $\tilde{G’}$ (squares), and loss, $\tilde{G''}$ (circles), moduli as a function of the reduced oscillatory frequency, $\tilde{\omega}$, for the viscoelastic samples, as indicated in the legend. Moduli, $G'$ and $G''$, are normalized by the sample shear modulus, $G_0$, and the frequency, $\omega$, is multiplied by the sample characteristic relaxation time, $\tau$. (Lines) Theoretical expectations for a one-mode Maxwell fluid.}
	\label{figure:2}
	\end{figure}

	For both classes of viscoelastic samples, the linear rheological behavior can be well described in the frequency range of interest by a one mode Maxwell model in which the viscoelasticity is characterized by a single relaxation time, $\tau$ and a plateau modulus, $G_\textrm{0}$. In this model, the storage and the loss moduli respectively read $G'(\omega)= \frac{G_\textrm{0}(\omega\tau)^2}{1+(\omega\tau)^2}$ and $G''(\omega)= \frac{G_\textrm{0}(\omega\tau)}{1+(\omega\tau)^2}$ with $\omega$, the oscillation frequency. In reduced units, the storage and the loss moduli respectively read $\tilde{G'}=G'/G_\textrm{0}=\frac{\tilde{\omega}^2}{1+\tilde{\omega}^2}$ and $\tilde{G''}=G''/G_\textrm{0}=\frac{\tilde{\omega}}{1+\tilde{\omega}^2}$ with $\tilde{\omega}=\omega\tau$. Fitting the frequency-dependent linear viscoelastic response of the samples with a Maxwell model allows one to determine the elastic plateau, $G_\textrm{0}$ and the characteristic time, $\tau$. These characteristic parameters are reported in table~\ref{t} along with the dynamic viscosity, $\eta_0= G_0 \tau$ for all viscoelastic samples. The plateau modulus, $G_0$, ranges from $10$ to $194$ Pa and the relaxation time, $\tau$, from $1$ ms to $178$ ms, yielding dynamic viscosities, $G_0 \tau$, between $0.031$ and $1.78$ Pa s. The samples are made to have relaxation times smaller than, comparable to, or larger than the characteristic time of the experiment, typically 6 ms, with the aim to probe viscous and elastic effects. We find (fig.~\ref{figure:2}) that the experimental data (symbols) for all viscoelastic samples collapse on one master curve which conforms to the one-mode Maxwell model (lines), showing that the samples viscoelasticity can indeed be satisfyingly described by this model.

\section{Experimental results}
	
\begin{figure}[!ht]
	
	\includegraphics[width=8.3 cm]{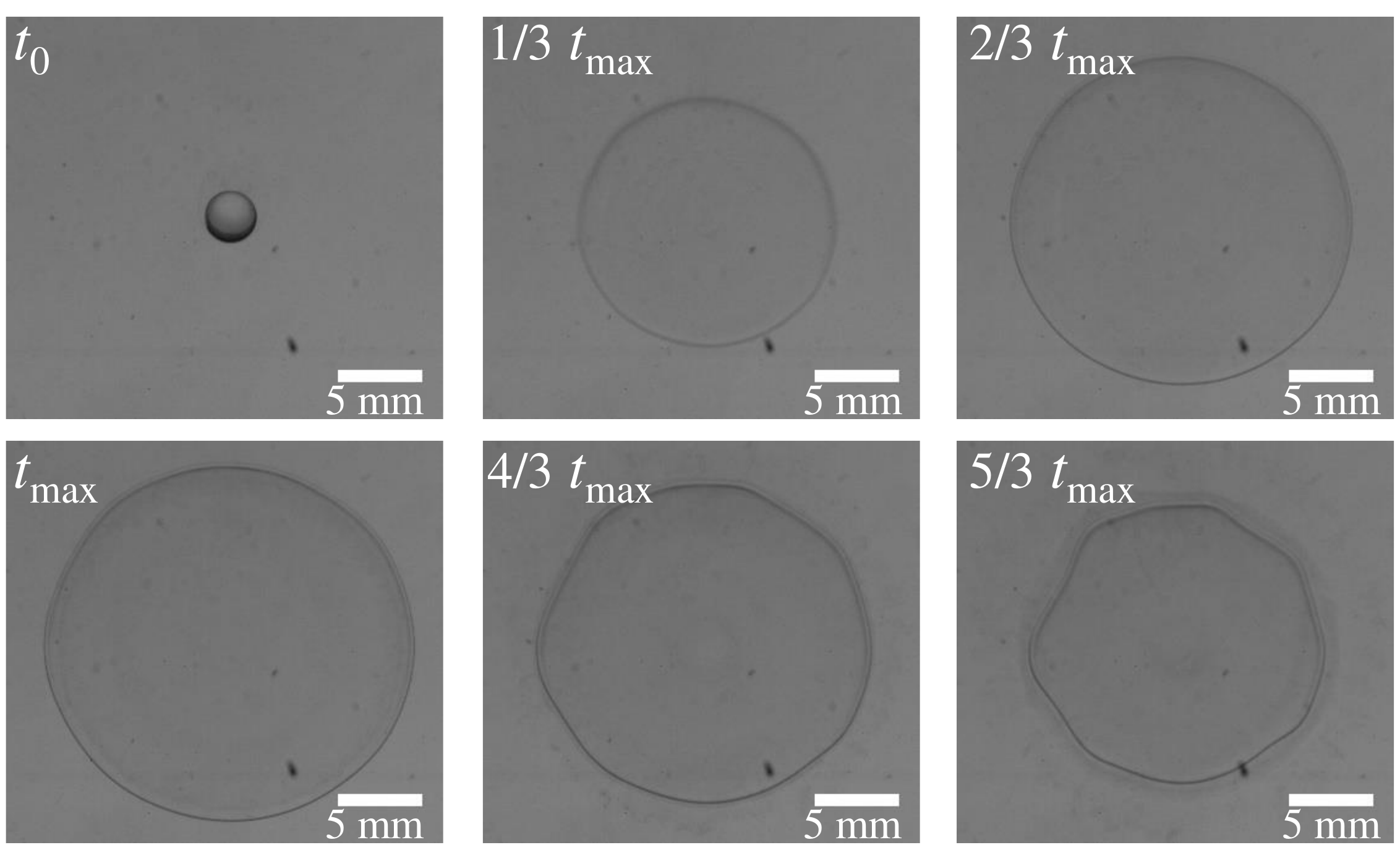}
	\caption{Snapshots taken at different times, as indicated, during the expansion and retraction of a sheet produced with sample M14$\phi$10\textit{r}6 impacted at a velocity of $4.7$ m/s. The bar sets the scale.}
	
	\label{figure:3}
	\end{figure}

\begin{figure}[!ht]
	
	\includegraphics[width=8.3 cm]{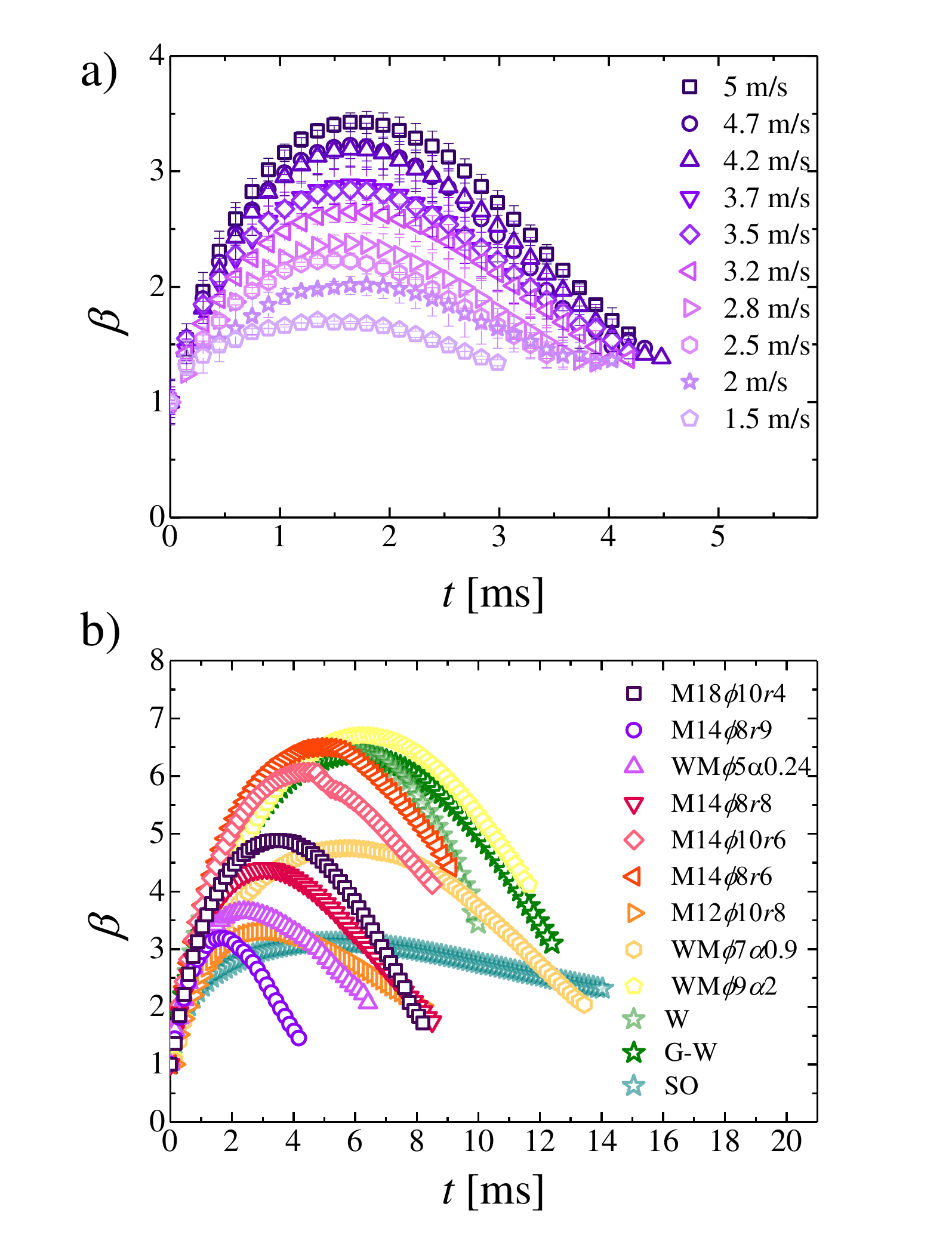}
	\caption{Time evolution of the sheet diameter upon impact normalized by the initial drop diameter for (a) sample M14$\phi$8\textit{r}9 at different impact velocities as indicated in the legend (b) all viscoelastic samples and selected Newtonian fluids: pure water (W, $\eta_0 = 1$ mPa s), glycerol water mixture (G-W, $\eta_0 = 216$ mPa s) and silicone oil (SO, $\eta_0 = 658$ mPa s) impacted at $v_0= 4.2$ m/s. The error bars in (a) represent the standard deviation of three different experiments. Error bars have been removed from (b) for more clarity.}
	
	\label{figure:4}
	\end{figure}

 After its release, the drop reaches the surface covered with a thin liquid nitrogen layer and expands with a disk-like shape on a gaseous nitrogen cushion. After reaching a maximal expansion, the sheet retracts due to surface tension and bulk elastic energy. A series of snapshots showing the whole process, from the impact of the drop to the expansion of the sheet and subsequently its retraction is displayed in figure~\ref{figure:3}. The expansion and retraction events result in bell shaped curves for the time evolution of the diameter of the sheet. During the retraction regime, axisymmetry is lost due to non reversible instabilities and the sheet cannot be reasonably assimilated to a disk anymore. For this reason, this regime will not be further investigated. We define $\beta=\frac{d}{d_0}$, the stretching ratio with $d$, the apparent diameter of the sheet and $d_0$, the initial diameter of the drop. The origin of time is chosen at the time at which the drop comes into contact with the nitrogen vapor layer in such a way that the diameter at $t=0$ is $d_0$, and hence $\beta= 1$. Figure~\ref{figure:4}a shows the time evolution of $\beta$ for sample M14$\phi$8\textit{r}9 at different impact velocities, $v_0$, from $1.5$ to $5$ m/s as indicated in the legend. The normalized diameter at maximum expansion, $\beta_{\textrm{max}}$, increases monotonically with the impact velocity while the time to reach maximal expansion, $t_{\textrm{max}}$, slightly increases with $v_0$. Figure~\ref{figure:4}b shows the time evolution of $\beta$ for all viscoelastic samples and three Newtonian samples with viscosity of $1$, $216$ and $658$ mPa s, for a same impact velocity $v_0 = 4.2$ m/s. Overall, the curves follow the same general behavior but the key parameters, $\beta_{\textrm{max}}$ and $t_{\textrm{max}}$, vary strongly from one sample to another (between $3.1$ and $6.7$ for $\beta_{\textrm{max}}$ and between $1.4$ and $6.3$ ms for $t_{\textrm{max}}$ at $v_0 = 4.2$ m/s) with no straightforward dependence on the rheological parameters, $G_0$, $\tau$ or $\eta_0$.

\begin{figure}[!ht]
	\includegraphics[width=8.3 cm]{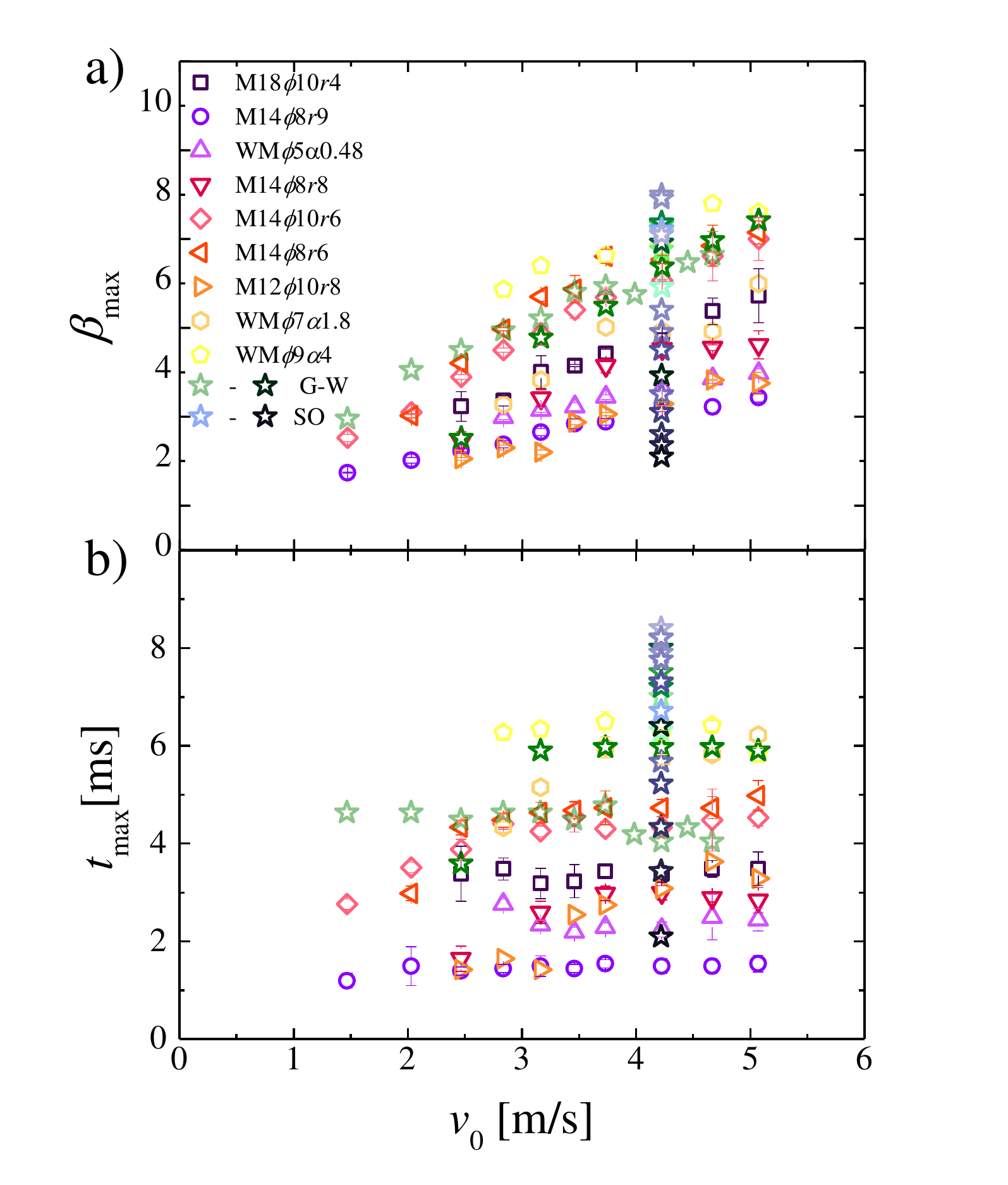}
	\caption{ Normalized maximal expansion (a) and time at maximal expansion (b) as a function of the impact velocity for all Newtonian and viscoelastic samples, as indicated in the legend. The increasing viscosity of the two classes of Newtonian samples are represented by shades of increasing darkness. The error bars represent the standard deviation of three different experiments.}
	\label{figure:5}
	\end{figure}

From the evolution of the diameter with time, we extract $\beta_{\textrm{max}}$ and $t_{\textrm{max}}$. The visoelastic fluids are impacted at different impact velocities and the dependence of $\beta_{\textrm{max}}$ and $t_{\textrm{max}}$ with the impact velocity is shown in figures~\ref{figure:5}a,b. Additionally, the effect of $v_0$ on $\beta_{\textrm{max}}$ and $t_{\textrm{max}}$ for two Newtonian samples: water and a mixture with $91$\% g/g of glycerol in water ($\eta_0= 216$ mPa s) is also shown in figures~\ref{figure:5}a,b. The other Newtonian samples (star symbols) are all impacted at one velocity ($v_0 = 4.2$ m/s) with $\beta_{\textrm{max}}$ and $t_{\textrm{max}}$ decreasing for increasing viscosity (symbols of increasing darkness). The dependence of $\beta_{\textrm{max}}$ with the impact velocity follows $\beta_{\textrm{max}}\propto v_0$ with prefactors which are sample-dependent. One could not infer simple correlations between the prefactors and the samples viscosity, bulk elasticity or relaxation time. On the other hand, $t_\textrm{max}$ is roughly constant with the impact velocity but varies from one sample to another. We observe a small decrease at low impact velocity for samples with relaxation time lower than the time of the experiment (M14$\phi$8\textit{r}8, M14$\phi$10\textit{r}6, M14$\phi$8\textit{r}6, M12$\phi$10\textit{r}8 and WM$\phi$7$\alpha$1.8).

	The next section of the paper aims to predict the experimental values obtained for $d_{\textrm{max}}$ and $t_{\textrm{max}}$ by rationalizing the interplay between viscosity, and surface and bulk elasticity, for the expansion dynamic.

\section{Damped oscillator model}
	
\subsection{Equation of motion}	

A common method to correlate the maximal stretching ratio, $\beta_{\textrm{max}}$, to the impact parameters is to balance inertia with the energy contributions arising either from surface tension~\cite{Richard2002,Biance2006,Andrew2017}, bulk elasticity~\cite{Tanaka2003,Tanaka2005} or both surface and bulk elasticity~\cite{Arora2018}. Below, we show that viscosity, surface tension and bulk elasticity are all crucial to quantitatively account for our experimental data. We write an energy balance equation for the expanding sheet after impact that includes at all times the contributions of the kinetic energy, $E_\textrm{K}$, the surface energy, $E_\gamma$, the bulk elastic energy, $E_{\textrm{bulk}}$, and the biaxial extensional viscous dissipation energy, $E_\textsc{b}$. 

\begin{equation}
E_\textrm{K}+E_\gamma+E_{\textrm{bulk}}+E_\textsc{b}=\textrm{Constant}
\label{balance}
\end{equation}

All energy terms are computed below considering the limit of large deformations ($d>>d_0$) and assuming that the sheet, at all times, is a disk of diameter, $d$, and of uniform thickness, $h_\textrm{s}$ \cite{Attane2007,Eggers2010}. From volume conservation, $h_\textrm{s}=\frac{2d_0^3}{3d^2}$. 
  
 The kinetic energy, $E_\textrm{K}$, reads

\begin{equation}
 E_\textrm{K}=\int_0^{d/2}\frac{1}{2}v_\textrm{r}^2\rho2\pi h_\textrm{s} r dr=\frac{m v^2}{4}
 \label{E_K}
\end{equation}

 with $m$, the mass of the drop, $d$, the diameter of the sheet at a time $t$ and $v_\textrm{r}=\frac{2vr}{d}$, the Eulerian velocity field in the material along $r$, the radial coordinate, where $v$ is the velocity of the expansion front. This velocity field has been found to describe accurately the expansion dynamics of Newtonian fluids~\cite{Vernay:2015dj,Raux:2020hs}.

The surface energy, $E_\gamma$ reads

\begin{equation}
E_\gamma=\frac{1}{2}\pi\gamma d^2
\label{E_gamma}
\end{equation}

The contribution of the disk edge to the surface energy is neglected due to the large deformation limit considered here.

The bulk elastic energy, $E_{\textrm{bulk}}$, is approximated by the biaxial extensional linear elastic deformation energy of a soft solid in the limit of large deformation~\cite{Tanaka2003,Arora2018}:

\begin{equation}
E_{\textrm{bulk}}\approx V_{\textrm{drop}}G_\textrm{eff}\left(\frac{d}{d_0}\right)^2
\label{E_bulk}
\end{equation}

with $V_{\textrm{drop}}=\frac{\pi d_0^3}{6}$, the volume of the drop and $G_\textrm{eff}$, the relevant shear elastic modulus of the sample, to be discussed in section \ref{param}. 

We assume that the unique source of dissipation is viscous and originates from the biaxial extensional deformation of the sheet \cite{Louhichi2020}. Indeed, dissipation due to shear is eliminated thanks to the non wetting and slip conditions achieved with the cold Leidenfrost effect leaving biaxial extensional viscous dissipation as the only remaining viscous contribution. The biaxial extensional viscous dissipation energy, $E_\textrm{B}$, reads: $E_\textrm{B}=\int_0^{t_{\textrm{max}}}\int_V\varphi dVdt$. Here, $\varphi$ is the dissipation function which reads $\varphi\approx\eta_\textrm{B,eff}\dot{\varepsilon}^2$ with $\dot{\varepsilon}=\frac{1}{d} \frac{\partial d}{\partial t}$, the Hencky strain rate and $\eta_\textrm{B,eff}$, the relevant biaxial extensional viscosity of the sample. In the following, we assume $\eta_\textrm{B,eff}$ to be constant with time for one given impact experiment (i.e. one given sample and one given impact velocity)~\cite{Louhichi2020} as discussed in section \ref{param}. 
	
	\begin{equation}
	E_\textrm{B}\approx\eta_\textrm{B,eff} \pi \frac{d_\textrm{0}^3}{6}\int_{0}^{t_{\textrm{max}}}\left(\frac{1}{d} \frac{\partial d}{\partial t}\right)^2dt
	\label{biaxial_dissip}
	\end{equation}

Inserting eqs.\ref{E_K}-\ref{biaxial_dissip} in the energy balance (eq.\ref{balance}) and deriving the latter with respect to time, one obtains the equation of motion for a free expanding sheet:

	  	 \begin{equation}
	\ddot{d}+\frac{c(d)}{m}\dot{d}+\omega_0^2 d=0
	\label{motion}
	\end{equation}

	Equation \ref{motion} is a non linear second order differential equation that can be viewed as the equation of motion of a damped harmonic oscillator with an angular frequency:
	\begin{equation}
	 \omega_0=\sqrt{\frac{ 8\pi}{m}\left(\gamma+\frac{d_\textrm{0}G_\textrm{eff}}{3}\right)}
	 \label{angular frequency}
	 \end{equation}
	  and a non constant viscous damping coefficient which decreases through the expansion, as $d$ increases.
	  
	  \begin{equation}
	 c(d)=\frac{4\pi d_\textrm{0}^3\eta_\textrm{B,eff}}{3}\frac{1}{d^2}
	 \label {damping}
	\end{equation}

In the absence of viscous dissipation ($c=0$) we retrieve the 1D undamped harmonic oscillator equation successfully used to model the expansion dynamics of sheets produced with various materials, with $\omega_0^2=\frac{8\pi\gamma}{m}$ for Newtonian fluids of low viscosity~\cite{Biance2003} (in that case $G_\textrm{eff}=0$), $\omega_0^2=\frac{8\pi d_0G_\textrm{eff}}{3m}$ for elastic beads~\cite{Tanaka2003} and $\omega_0^2=\frac{ 8\pi}{m}\left(\gamma+\frac{d_\textrm{0}G_\textrm{eff}}{3}\right)$ for ultrasoft elastic beads~\cite{Arora2018}. When viscous dissipation is not negligible, the relevant rheological parameters to consider are less straightforward, and are discussed below.

\subsection{Discussion on the relevant rheological parameters}	
\label{param}
\begin{figure}[H]
	\centering
	\includegraphics[width=8.3 cm]{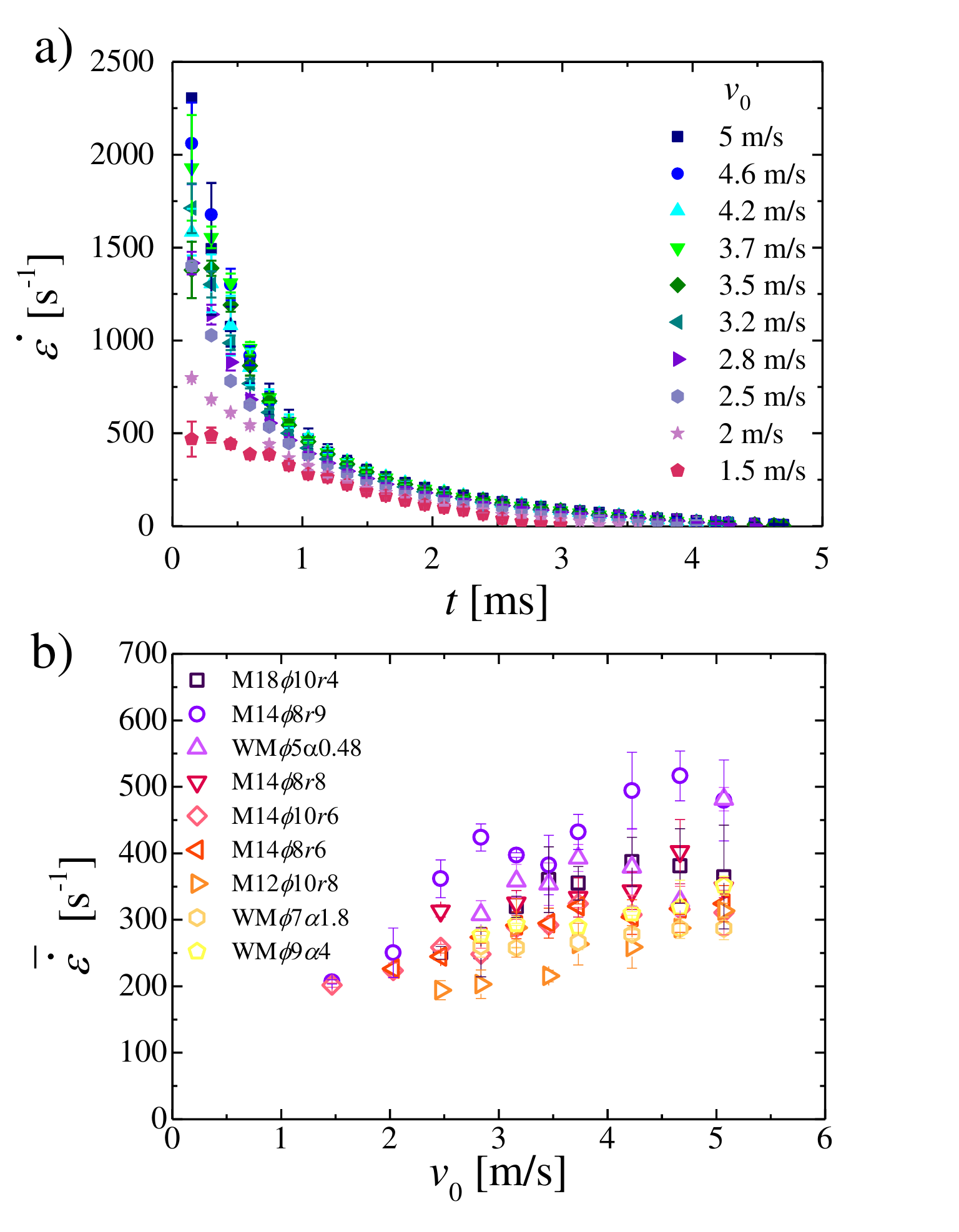}
	
	\caption{(a) Time evolution of the Hencky strain rate during the sheet expansion for the sample M14$\phi$8\textit{r}9 at different impact velocities, $v_0$, as indicated in the legend. (b) Strain rate averaged over time as a function of $v_0$ for all viscoelastic samples. Error bars represent the standard deviation of three different experiments.}
	\label{figure:6}
	\end{figure}

When viscous dissipation is not negligible, $c\neq0$ is proportional to the biaxial extensional viscosity $\eta_\textrm{B,eff}$.
In the case of Newtonian fluids ($G_\textrm{eff}=0$), $\eta_\textrm{B,eff}=6\eta_0$, the zero-shear viscosity multiplied by the Trouton ratio~\cite{Macosko:1994bk,Louhichi2020}. The Newtonian sample biaxial extensional  viscosities thus range from $6$ to $4880$ mPa s for mixtures of glycerol and water and from $30$ to $5820$ mPa s for silicone oil blends.
For the Maxwell fluids, the approach is less straightforward as the biaxial extensional  viscosity, but also the elastic modulus, are expected to be time-dependent to account for the samples viscoelasticity~\cite{Arora2016}. Hence, we need to estimate $G_\textrm{eff}$ and $\eta_\textrm{B,eff}$ at the relevant frequency. As a first order approach, the relevant frequency can be estimated as the Hencky strain rate, $\dot{\varepsilon}$, undergone by the sheet during its expansion. The Hencky strain rate is non-stationary and vanishes at maximal expansion as illustrated in figure~\ref{figure:6}a for one sample (M14$\phi$8\textit{r}9) impacted at different velocities. We choose as characteristic frequency the mean value of $\dot{\varepsilon}$ averaged over the duration of the  expansion regime: $\bar{\dot{\varepsilon}}=\frac{\int_0^{t_{\textrm{max}}}\dot{\varepsilon}\ dt}{t_{\textrm{max}}}$. The mean values are plotted in figure~\ref{figure:6}b as a function of the impact velocity for all viscoelastic samples. We measure that $\bar{\dot{\varepsilon}}$ ranges between $194$ and $516$ s$^{-1}$ corresponding to Weissenberg number, Wi$=\bar{\dot{\varepsilon}}\tau$, values from 0.3 to 68.9. We find that $\bar{\dot{\varepsilon}}$ varies almost linearly with $v_0$, with a sample-dependent proportionality constant. In the following, for Maxwell fluids, the relevant shear elastic modulus, $G_\textrm{eff}$, is taken as the elastic part, $G'$, of the complex modulus, $G^*$, at the relevant frequency:

  \begin{equation}
  G_\textrm{eff}=G'({\bar{\dot{\varepsilon}}})=G_0\frac{(\tau \bar{\dot{\varepsilon}})^2}{1+(\tau \bar{\dot{\varepsilon}})^2}
 \end{equation}

  Similarly, the biaxial extensional viscosity for a viscoelastic fluids at the relevant strain rate is evaluated using the viscous part, $G''$, of the complex modulus, $G^*$, and assuming the same Trouton ratio as for the Newtonian fluids: 
	
	\begin{equation}
	\eta_\textrm{B,eff}=6\frac{G''(\overline{\dot{\varepsilon}})}{\overline{\dot{\varepsilon}}}=6G_0\frac{\tau}{1+(\tau \bar{\dot{\varepsilon}})^2}
	\end{equation}

\subsection{Approximate analytical solution and comparison with experiments}	

\begin{figure}[H]
	\centering	
	\includegraphics[width=8.3 cm]{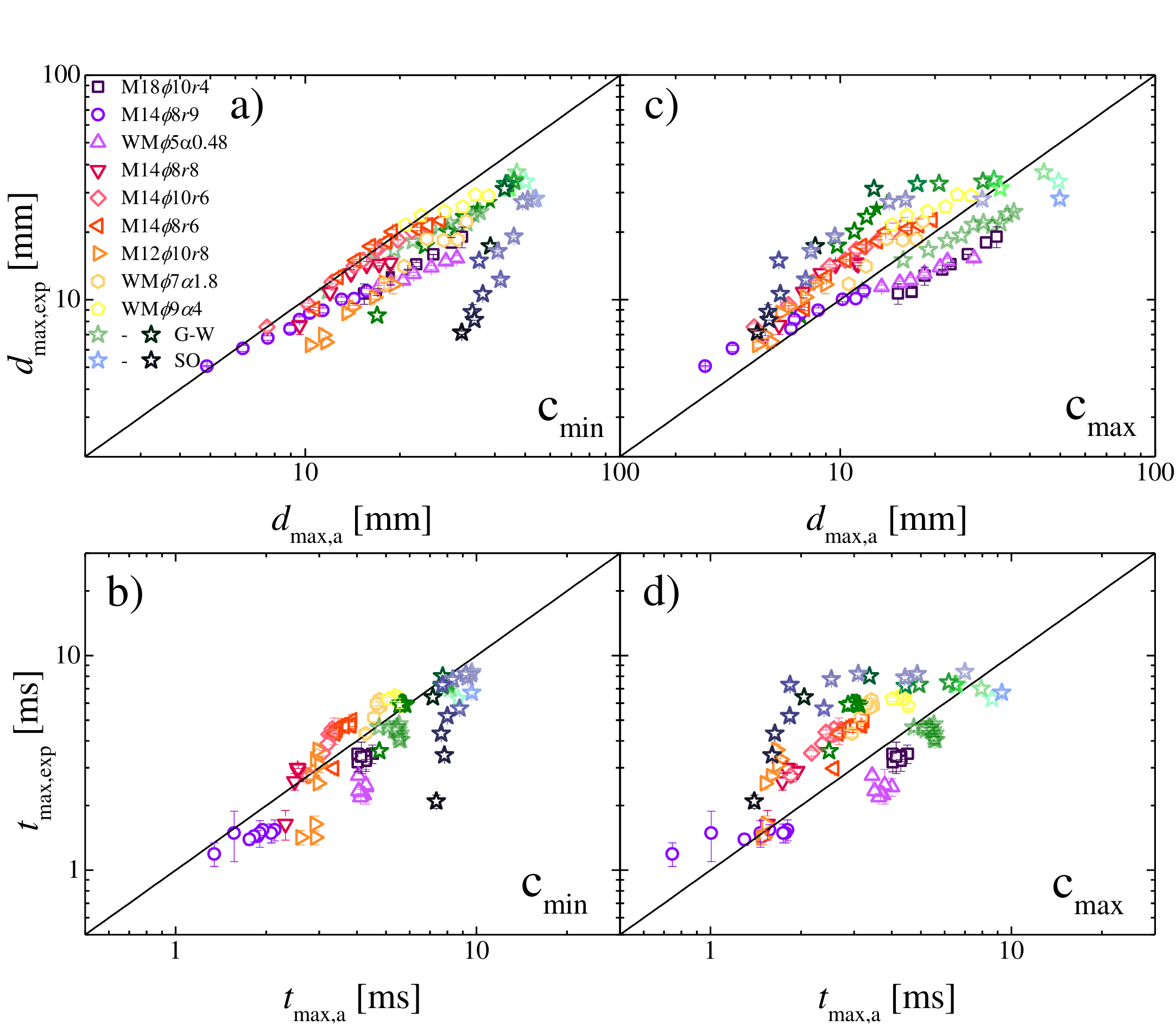}
	
	\caption{  Experimental $\textit{d}_{\textrm{max}}$ (a, c) and $\textit{t}_{\textrm{max}}$ (b, d) as a function of their respective theoretical predictions using the damped harmonic oscillator assumption with an underestimation ($c_{\textrm{min}}$, a, b) or overestimation ($c_{\textrm{max}}$, c, d) of the damping coefficient for all samples. The lines represent x=y.}
	\label{figure:7}
	\end{figure}

	Equation \ref{motion} is not analytically solvable. Therefore, we propose
 to solve the problem analytically for two limits of the damping coefficient:  (i) $c\approx c_{\textrm{min}}=c(d_{\mathrm{max}})=\frac{4\eta_\textrm{B}\pi d_0^3}{3 d_{\textrm{max}}^2}$ and (ii) $c\approx c_{\textrm{max}}=c(d_0)=\frac{4\eta_\textrm{B}\pi d_0}{3}$ . In these two cases, the equation of motion of the free sheet (eq.\ref{motion}) reduces to a simple damped harmonic oscillator model with a constant damping factor. Using $c_{\textrm{max}}$, we overestimate the effect of the viscous drag and using $c_{\textrm{min}}$, we underestimate it. We check that all our samples verify the condition $\frac{c}{2m}<\omega_0^2$ which corresponds to the underdamped regime of the harmonic oscillator. In this regime, the solution of equation \ref{motion} is
\begin{equation}
d(t)=C e^{-\frac{c}{2m}t} \cos(\omega_\textrm{d}t-\Phi)
\end{equation}
The parameters $\omega_\textrm{d}=\sqrt{\omega_0^2-\left(\frac{c}{2m}\right)^2}$, $C=d_\textrm{0}\sqrt{\left(\frac{2v_\textrm{0}}{d_\textrm{0}\omega_\textrm{d}}+\frac{c}{2m\omega_\textrm{d}}\right)^2+1}$ and $\Phi=\tan^{-1}\left(\frac{2v_\textrm{0}}{d_\textrm{0}\omega_\textrm{d}}+\frac{c}{2m\omega_\textrm{d}}\right)$ are obtained from the initial conditions, $d(0)=d_0$ and $\dot{d}(0)=2v_0$ (See supplementary materials \dag).\\
 The maximum of $d$ gives the theoretical time at maximal expansion which allows one to obtain the analytic equations, $t_{\textrm{max,a}}$ and $d_{\textrm{max,a}}$.
\begin{equation}
t_{\textrm{max,a}}=\frac{1}{\omega_d}\arctan\left(\frac{2v_0\omega_d }{d_0\omega_d^2+\frac{c v_0 }{m}+\frac{d_0c^2}{4m^2}}\right)
\label{t_max_a}
\end{equation}
\begin{equation}
d_\textrm{max,a}=A e^{-\frac{c}{2m}t_\textrm{max,a}} \cos(\omega_\textrm{d}t_\textrm{max,a}-\Phi)
\label{d_max_a}
\end{equation}
In order to validate the present model, the analytical approximate predictions of the maximal diameter, $d_\textrm{max,a}$ and the time at maximal expansion, $t_\textrm{max,a}$ are compared with the experimental results. Figure~\ref{figure:7} shows the experimental values for $d_{\textrm{max}}$ and $t_{\textrm{max}}$
plotted against the theoretical ones when using the relevant parameters of the drops in the underdamped harmonic oscillator model approximation (eqs.\ref{t_max_a}, \ref{d_max_a}). In figure~\ref{figure:7}(a,b) the dissipation is evaluated with $c=c_{\textrm{min}}$ and in figure~\ref{figure:7}(e, d ) with $c=c_{\textrm{max}}$. Lines representing x=y allow one to judge more easily the quality of the analytical approximation of the model. The agreement between the underdamped harmonic oscillator model and the experimental data is good (relative error of approximately 30$\%$). As expected, when the viscous dissipation is underestimated ($c=c_{\textrm{min}}$), respectively overestimated ($c=c_{\textrm{max}}$), overall the analytical predictions are larger (respectively smaller) than the experimental values and the data points are all below (respectively above) the lines $d_{\textrm{max,a}}=d_{\textrm{max,exp}}$ and $t_{\textrm{max,a}}=t_{\textrm{max,exp}}$.\\

\subsection{Numerical resolution and comparison with experiments}

\begin{figure}[h]
	\centering
	\includegraphics[width=8.3 cm]{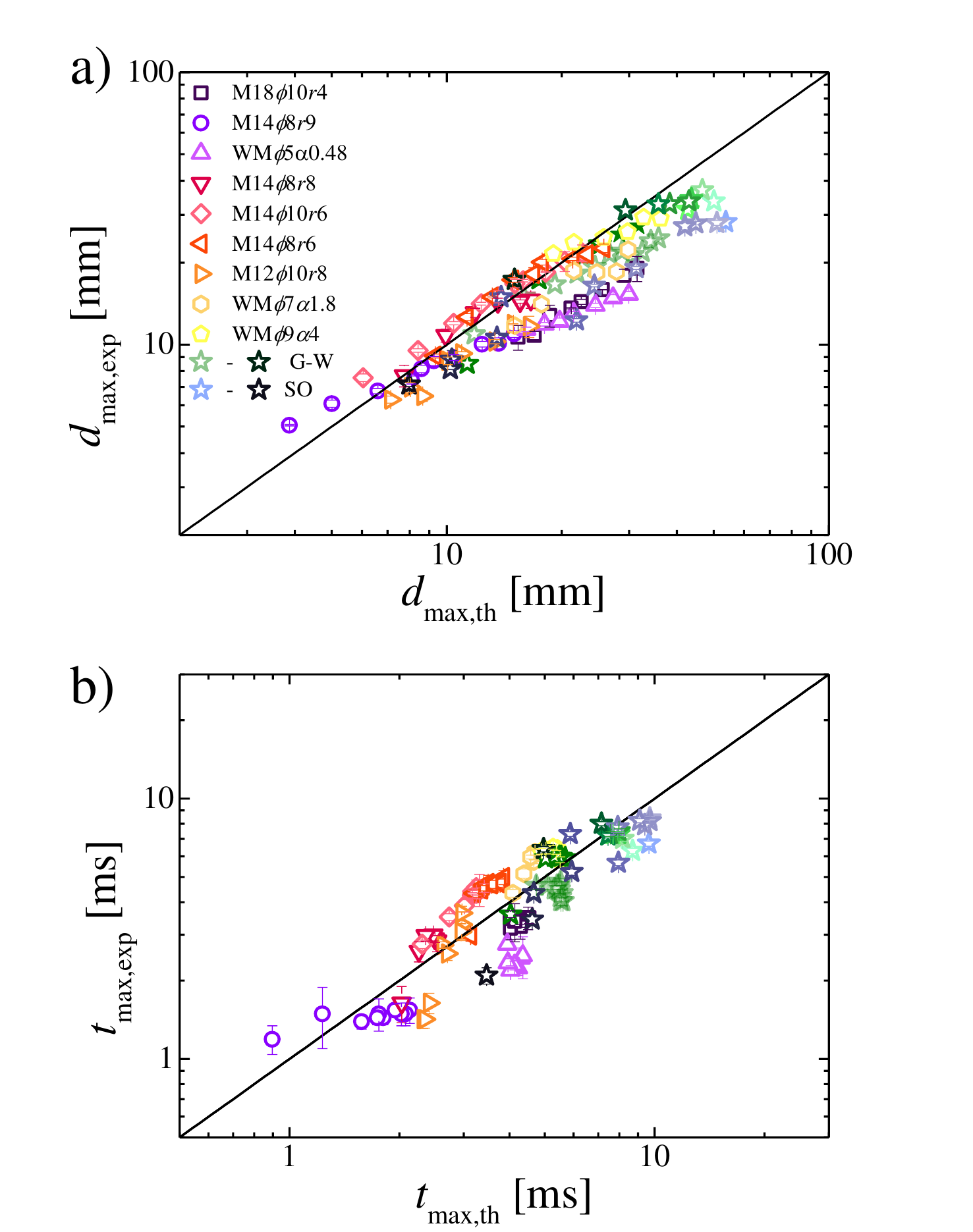}
	
	\caption{Experimental (a) $\textit{d}_{\textrm{\textrm{\textrm{max}}}}$, and (b) $\textit{t}_{\textrm{\textrm{\textrm{max}}}}$ as a function of their respective theoretical predictions using numerical resolution of equation \ref{motion} for all samples. The lines represent x=y. }
	\label{figure:8}
	\end{figure}

To refine the prediction and better account for the competition between non-stationary elasticity, viscous dissipation and capillarity, we solve equation \ref{motion} numerically. Figure~\ref{figure:8} shows the experimental data against the theoretical predictions after numerical resolution. The results agree well with the prediction for $d_{\textrm{max}}$ as well as for $t_{\textrm{max}}$ considering that no adjustable parameters are used to determine theoretical predictions. We can predict $d_{\textrm{max}}$ and $t_{\textrm{max}}$ with a relative error of approximately 25$\%$ for Newtonian fluids with biaxial extensional viscosities varying over three order of magnitude and viscoelastic samples spanning a large range of elastic moduli, characteristic relaxation times and viscosities. Despite the fact that we investigate Newtonian samples of high viscosity and viscoelastic fluids, we find comparable relative error as previous works focusing on low viscosities Newtonian fluids impacted in similar conditions. Indeed, a relative error of 20$\%$, at best, has been found when comparing experimental maximal expansion of Newtonian fluids of low vicosities impacted on a repellent surface\cite{Pearson2012} to theoretical predictions using analytical, empirical, and scale relations\cite{Clanet2004,Roisman2009} as well as dynamical models\cite{Pasandideh-Fard1996,Roisman2002,Ukiwe2005}.  

\begin{figure}[h]
	\centering
	\includegraphics[width=8.3 cm]{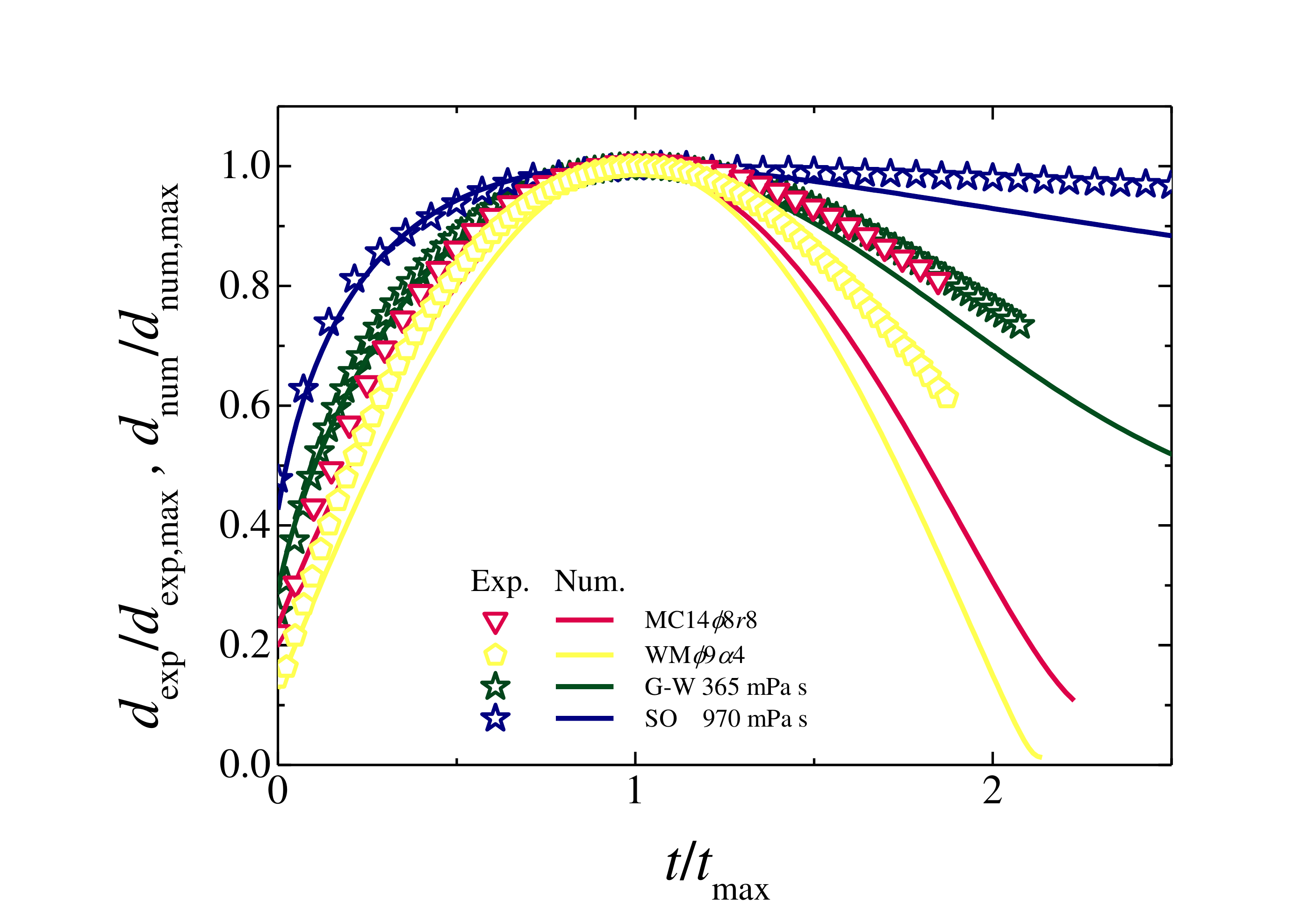}
	
	\caption{ Experimental (symbols) and theoretical (eq. \ref{motion}) (lines) evolution of the diameter of the sheet normalized by its value at maximal expansion as a function of the time normalized by the time at maximal expansion for samples M14$\phi$8\textit{r}8 and WM$\phi$9$\alpha$4, a glycerol water mixture (G-W, $\eta_0 = 365$ mPa s) and a silicone oil (SO, $\eta_0 = 970$ mPa s) impacted at $v_0$ = 4.2 m/s.}
	\label{figure:9}
	\end{figure}

Furthermore, the dynamics of expansion of the sheets can be relatively well described from only two parameters, $t_{\textrm{max}}$ and $d_{\textrm{max}}$. Figure~\ref{figure:9} shows, for four representative samples, the evolution of the experimental diameter (symbols) along with its numerical prediction using equation \ref{motion} (lines) normalized by their maximal values, as a function of the time normalized by $t_\textrm{max}$. In the expansion regime, a maximal error of 5$\%$ is found between the normalized experimental data and the corresponding normalized theoretical predictions. Much larger deviations can be observed in the receding regime as different complex phenomena such as the loss of axisymmetry, expulsion of secondary droplets or pinning due to the cold surface, which are not taken into account in our rationalization, could occur. 
	
	 \section {Conclusion}

Drop impact experiments on repellent surfaces have been performed with Newtonian and Maxwell fluids. Maxwell fluids have been carefully chosen to display a wide range of relaxation times, shorter, comparable or much larger than the typical experimental time, allowing us to investigate the roles of viscosity, bulk elasticity and capillarity in the expansion dynamics of fluid sheets. We have used a cold Leidenfrost effect-based set-up so that the dominant source of viscous dissipation is the biaxial extensional deformation of the sheet. We have provided a systematic study of the effect of the impact velocity on the maximum diameter of the expanding sheet, and the time at which maximum expansion is reached.
The  expansion dynamics could be successfully modeled by a non linear damped harmonic  oscillator model, where the damping coefficient decreases during the expansion and is proportional to the  biaxial extensional dynamic viscosity, and the undamped angular frequency of the oscillator results from a simple combination of the surface tension and the dynamic modulus of the sample. For the viscoelastic samples, we have proposed to take as dynamic viscosity and dynamic modulus the viscosity and elastic modulus at a characteristic rate equal to the mean Hencky  strain rate of the sheet in the expansion regime.  The numerical prediction for the maximal expansion diameter, $d_{\textrm{max}}$, and the time to reach maximal expansion, $t_{\textrm{max}}$, agree quantitatively well with the experimental results, without adjustable parameters. Our approach is simple but quite general and we believe it could successfully be applied to more complex viscoelastic materials, of relevance for applications.\\

\section*{Conflicts of interest}
	
There are no conflicts to declare.

\section*{Acknowledgements}

	This work was financially supported by the H2020 Program (Marie Curie Actions) of the European Commission's Innovative Training Networks (ITN) (H2020-MSCA-ITN-2017) under DoDyNet REA Grant Agreement (GA) $N^\circ.765811$. We thank Pr. Daniel Read (University of Leeds) and Pr. Dimitris Vlassopoulos (IESL-FORTH) for fruitful discussions and Dr. Ty Phou (L2C, Montpellier) for his technical assistance.



\balance


\bibliography{library} 
\bibliographystyle{rsc} 

\end{document}